\documentclass{ifacconf}

\usepackage{graphicx}      
\usepackage{natbib}        
\usepackage{amsmath}
\usepackage{amssymb}
\usepackage{xcolor}
\usepackage{siunitx}
\usepackage{algorithm}
\usepackage{algpseudocode}
\usepackage{subfigure}
\usepackage{float}
\theoremstyle{definition}

\usepackage{mathtools}

\DeclareMathOperator*{\minimize}{minimize}

\newtheorem{assumption}{Assumption}

\begin{document}
\begin{frontmatter}

\title{Multi-Layer Continuum Deformation Optimization of Multi-Agent Systems}

\thanks[footnoteinfo]{This work has been supported by the National Science Foundation under Award Nos. 2133690 and 1914581.}

\author[First]{Harshvardhan Uppaluru} 
\author[Second]{Hossein Rastgoftar} 

\address[First]{Scalable Move And Resilient Traversability Lab, Aerospace and Mechanical Engineering department, University of Arizona, Tucson - 85721, USA. {Email: huppaluru@arizona.edu}}
\address[Second]{Scalable Move And Resilient Traversability Lab, Aerospace and Mechanical Engineering department, University of Arizona, Tucson - 85721, USA. {Email: hrastgoftar@arizona.edu}}

\begin{abstract}
This paper studies the problem of safe and optimal continuum deformation of a large-scale multi-agent system (MAS). We present a novel approach for MAS continuum deformation coordination that aims to achieve safe and efficient agent movement using a leader-follower multi-layer hierarchical optimization framework with a single input layer, multiple hidden layers, and a single output layer. 
The input layer receives the reference (material) positions of the primary leaders, the hidden layers compute the desired positions of the interior leader agents and followers, and the output layer computes the nominal position of the MAS configuration. By introducing a lower bound on the major principles of the strain field of the MAS deformation, we obtain linear inequality safety constraints and ensure inter-agent collision avoidance. The continuum deformation optimization is formulated as a quadratic programming problem. It consists of the following components: (i) decision variables that represent the weights in the first hidden layer; (ii) a quadratic cost function that penalizes deviation of the nominal MAS trajectory from the desired MAS trajectory; and (iii) inequality safety constraints that ensure inter-agent collision avoidance. To validate the proposed approach, we simulate and present the results of continuum deformation on a large-scale quadcopter team tracking a desired helix trajectory, demonstrating improvements in safety and efficiency.

\end{abstract}

\begin{keyword}
 Multi-agent systems, Flying robots, Networked robotic system modeling and control, Continuum Mechanics
\end{keyword}

\end{frontmatter}
\begin{figure}[ht]
    \centering
    \includegraphics[width=\linewidth]{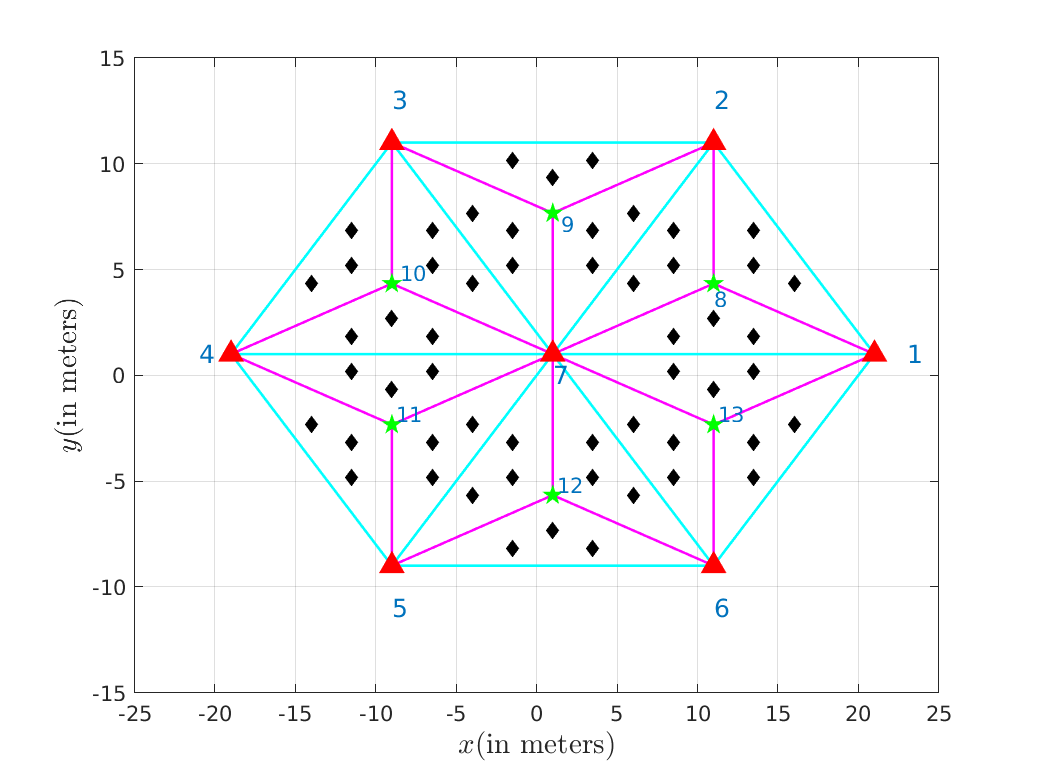}
     \vspace{-0.5cm}
    \caption{An example of the configuration of a $N = 67$ agent team defined by set $ \mathcal{V}=\{1,\cdots,67\}$, where $\mathcal{V}=\mathcal{W}_1\bigcup \mathcal{W}_2\bigcup \mathcal{W}_3$. The set $\mathcal{W}_1 = \{1,\cdots,7\}$ is represented by red triangles where $1$ through $6$ are the boundary leader agents and $7$ is the core leader agent. Set $\mathcal{W}_2=\mathcal{W}_1 \cup \{8,\cdots,13\}$, set $\mathcal{W}_3=\mathcal{W}_2 \cup \{14,\cdots,67\}$ identifies all the agents in the $N$-agent configuration. We also define the set $\mathcal{T} = \{1,\cdots,6\}$ that identifies the number of unique triangles formed by the leading polygon. Set $\mathcal{V}_j$ associates particular agents enclosed by triangle $j \in \mathcal{T}$.}
    \label{fig:configuration_plot}
\end{figure}

\section{Introduction}
\label{sec:introduction}
Coordination, formation control and cooperative control are important active areas of research in the field of Multi-Agent Systems(MAS) with a wide range of potential applications including, but not limited to, surveillance \citep{leslie2022unmanned}, search and rescue operations \citep{kleiner2013rmasbench}, and air traffic monitoring \citep{idris2018air}. In terms of efficiency, costs, and resilience to failure, a MAS comprised of Unmanned Aerial Vehicles (UAV) can provide substantial benefits over a single UAV. Cooperation among MAS agents improves the multi-agent team's capacity to recover from abnormalities. Virtual structure, containment control, consensus control and continuum deformation \citep{rastgoftar2016continuum} are some of the prominent and existing approaches for multi-agent system coordination that have been extensively studied. 

\subsection{Related Work}
The centralized coordination approach in which the multi-agent formation is represented as a single structure and a rigid body is known as Virtual Structure \citep{lewis1997high, beard2000feedback}. Given some orientation in $3$-D motion space, the virtual structure moves in a certain direction as a rigid body while maintaining the rigid geometric relationship between multiple agents. 
Consensus control \citep{cao2015leader, shao2018leader} is a decentralized technique with various coordination implementations presented such as leaderless multi-agent consensus \citep{ding2019leaderless,qin2016leaderless} and leader-follower consensus \citep{wu2018leader}.
Another decentralized leader-follower technique is containment control \citep{notarstefano2011containment}, where a fixed number of leaders guide the followers via local communication. A multi-agent system's finite-time containment control has been investigated \citep{wang2013distributed, liu2015distributed}. The criteria required and adequate for containment control stability and convergence were developed in \citep{cao2012distributed, ji2008containment}. Researchers investigated containment control in the context of fixed and switching inter-agent communication \citep{li2015containment}. 

Continuum deformation \citep{rastgoftar2016continuum, rastgoftar2017continuum, UPPALURU2022107960} is a decentralized multi-agent coordination technique that considers agents as finite number of particles in a continuum that deform and translate in $3$-D space. An $n$-D $(n = 1, 2, 3)$ continuum deformation coordination has at least $n+1$ leaders in $\mathbb{R}^n$, positioned at the vertices of an $n$-D simplex at any time $t$. Leaders plan the agent team's continuum deformation independently which are  acquired by followers  through local communication. Despite the fact that both containment control and continuum deformation are decentralized leader-follower approaches, continuum deformation enables inter-agent collision avoidance, obstacle collision avoidance, and agent containment by formally specifying and verifying safety in a large-scale multi-agent coordination \citep{rastgoftar2018safe, rastgoftar2019safe}. A large scale multi-agent system can safely and aggressively deform in an obstacle-filled environment by employing continuum deformation coordination. Experimental evaluation of continuum deformation coordination in $2$-D with a team of quadcopters has been performed previously \citep{romano2019experimental, romano2022quadrotor, uppaluru2022drones}.
\subsection{Contributions and Outline}
We advance the existing continuum deformation approach \citep{rastgoftar2016continuum, rastgoftar2017continuum, UPPALURU2022107960} towards multi-layer continuum deformation (MLCD) coordination at which the desired multi-agent deformation is planned by a finite number of leaders organized hierarchically through a feed-forward network. The feed-forward hierarchical network consists of one input layer receiving reference positions of the boundary agents, $p$ hidden layers, and one output layer. While the first $p-1$ hidden layers contain neurons that represent leaders, the neurons sorted in the last hidden layer (hidden layer $p$) all represent follower agents. The  output layer computes  nominal position of the agent team configuration by minimizing the error between nominal  and desired trajectories. The proposed MLCD overcomes the deformation uniformity of the available continuum deformation coordination, which is resulted from deformation planning by a single Jacobian matrix.  MLCD is defined as a quadtratic programming problem with inequality safety constraints obtained by eigen-decomposition of MAS spatial deformation matrices.

The rest of the paper is organized as follows: Preliminaries are first introduced in Section \ref{sec:preliminaries} followed by a detailed description of our approach in Section \ref{sec:proposedapproach}. Safety guarantee conditions have been presented in Section \ref{sec:safetyconditions} before presenting simulation results in Section \ref{sec:simulations}. Section \ref{sec:conclusion} concludes the paper. 
\begin{figure}[ht]
    \centering
    \includegraphics[width=0.48\textwidth]{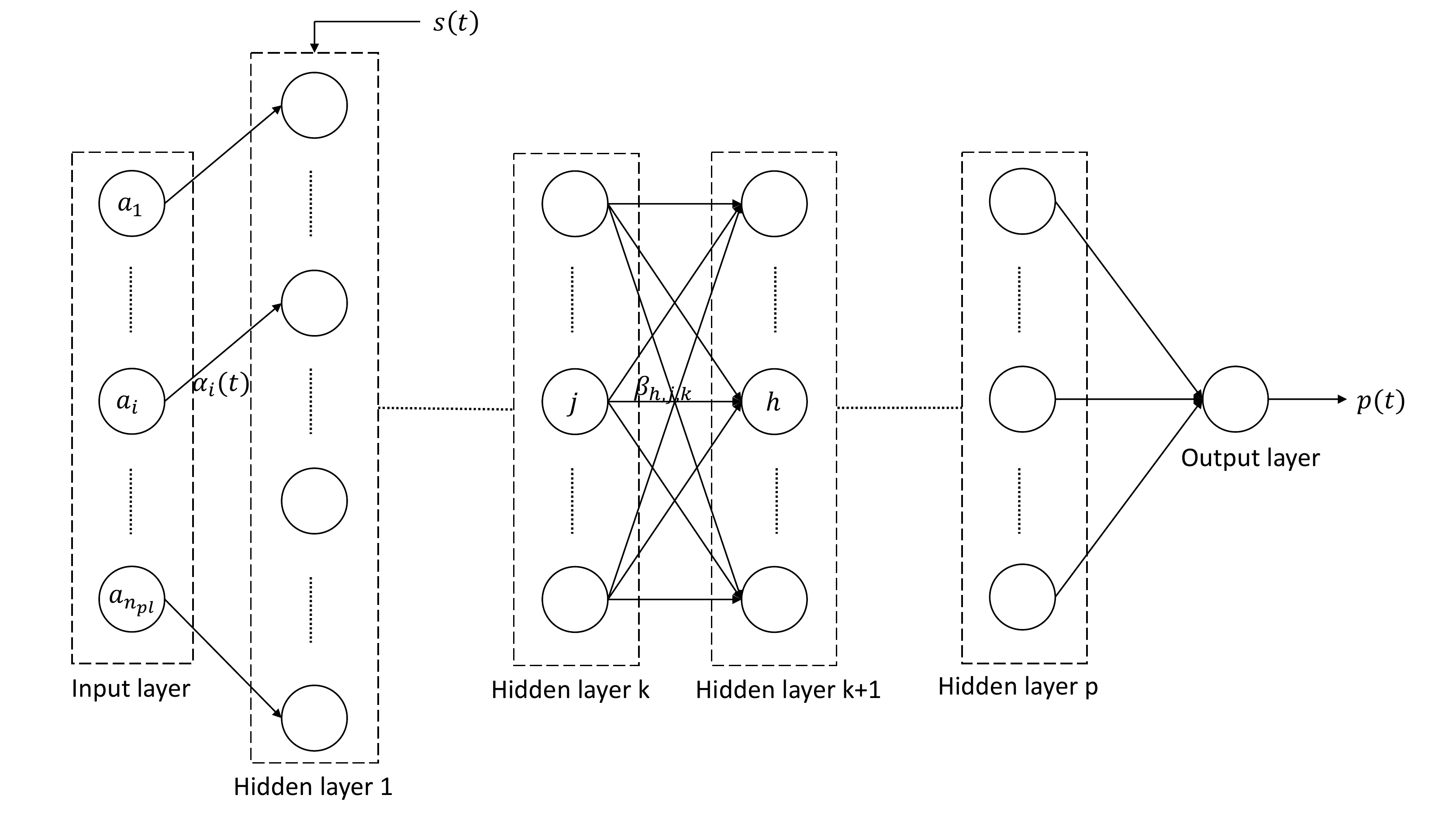}
     \vspace{-0.5cm}
    \caption{The structure of the proposed hierarchical optimization approach inspired from neural networks.}
    \label{fig:nn}
\end{figure}
\section{Preliminaries}
\label{sec:preliminaries}

We define a $N$-agent team as individual particles of a deforming body navigating collectively in a $3$-D motion space. An example of the $N$-agent team configuration where $N = 67$ is shown in Figure \ref{fig:configuration_plot}. The set $\mathcal{V}=\left\{ 1,\cdots, N\right\}$ is used to uniquely identify all the agents in the team. 
Set $\mathcal{V}$ can be expressed as
\begin{equation}
    \centering
    \label{eq:perceptron_based}
    \mathcal{V} = \bigcup_{k=1}^p\mathcal{W}_k
\end{equation}
with subsets $\mathcal{W}_1$ through $\mathcal{W}_p$, where $\mathcal{W}_1$ defines the boundary leader agents and a single core (leader) agent located inside the agent team configuration; $\mathcal{W}_2$ through $\mathcal{W}_{p-1}$ define the interior leaders; and $\mathcal{W}_p$ defines pure follower agents.  
The boundary and core agents defined by $\mathcal{W}_1$ are called \textit{primary leaders} and $n_{pl}=\left|\mathcal{W}_1\right|$ denotes the number of primary leaders.

Subset $\mathcal{W}_k$ functions as immediate leaders for $\mathcal{W}_{k+1}$ (for $k=1,\cdots,p-1$), which implies that desired positions of the agents belonging to $\mathcal{W}_{k+1}$ are obtained based upon the desired positions of the agents in $\mathcal{W}_k$. 
Explicitly, the desired position of agent $i \in \mathcal{W}_{k+1}$ is given by
\begin{equation}
    \centering
    \label{eq:desired_positions_agents}
    \mathbf{p}_i(t)=\sum_{j\in \mathcal{W}_k}\beta_{i,j,k}\mathbf{p}_j(t),\quad ~k=2,\cdots,p-1,
\end{equation}
where $\beta_{i,j,k}\in \left[0,1\right]$, and
\begin{equation}
    \centering
    \sum_{j\in \mathcal{W}_k}\beta_{i,j,k}=1,\qquad i\in \mathcal{W}_{k+1},~k=2,\cdots,p-1.
\end{equation}
The desired trajectories of agents belonging to set $\mathcal{W}_1$ are calculated using
\begin{equation}
    \centering
    \label{eq:desired_leaders_positions}
    \mathbf{p}_l(t)=\alpha_{l}(t)\mathbf{a}_{l0}+\mathbf{s}(t),\qquad \forall l\in \mathcal{W}_1,
\end{equation}
where $\alpha_{l}(t)$ is a positive weight parameter $\in \left[ \alpha_{min}, \alpha_{max} \right]$, $\mathbf{a}_{l0}$ specifies the constant reference (material) position of agents $l \in \mathcal{W}_1$ at time $t_0$, $\mathbf{s}(t)$ is the desired position of the $N$-agent team configuration with respect to an inertial coordinate system. Note that the reference position of the core agent $l \in \mathcal{W}_1$ is $\mathbf{0}$, i.e. $\mathbf{a}_{n_{pl}}=\mathbf{0}$.

\begin{assumption}\label{leadermagassume}
We assume that the magnitude of reference position of every boundary leader $l\in \mathcal{W}_1\setminus \left\{n_{pl}\right\}$ is the same and it is equal to $a_0$. Therefore,

\begin{equation}
    \bigwedge_{l\in \mathcal{W}_1\setminus \left\{n_{pl}\right\}}\left(\|\mathbf{a}_{l0}\|=a_0\right).
\end{equation}
\end{assumption}
\section{Proposed Approach}
\label{sec:proposedapproach}
We propose a hierarchical optimization framework for safe and optimal continuum deformation coordination of a $N$-agent team over the finite time interval $[t_0, t_f]$. The proposed hierarchical structure resembles a Neural Network (See Figure. \ref{fig:nn}) and consists of one input layer, hidden layers, and a single output layer, where the nodes contained by non-output layers represent agents. The hierarchical framework is structured as follows:


As shown in Figure. \ref{fig:nn}, the input layer receives the time-invariant reference (material) position $(\mathbf{a}_{l0})$ of primary leader agents at time $t_0$.
The first hidden layer computes the desired trajectories of boundary and core agents using equation (\ref{eq:desired_leaders_positions}).
More specifically, in hidden layer $1$, node $j \in \mathcal{W}_1$,  receives reference position $\mathbf{a}_{j0}$, and bias $\mathbf{s}(t)$, to return $\mathbf{p}_j(t)$ according to equation (\ref{eq:desired_leaders_positions}). 
The weights associated with this hidden layer are coined as $\boldsymbol{\alpha}$(t). Every hidden layer $k \in \{2,\cdots,p\}$ acquires desired positions from previous hidden layer $k-1$ and yields the desired positions of the agents defined by $\mathcal{W}_k$ by
using Eq. (\ref{eq:desired_positions_agents}). Note that bias is $\mathbf{0}$ in these hidden layers $2$ through $p$. The weights associated with hidden layer $k \in \{2,\cdots,p\}$ is coined as $\boldsymbol{\beta}_k$ and are constant. 

The output layer generates the nominal position of $N$-agent team configuration which is denoted by $\mathbf{p}(t)$ by averaging the desired positions of agents in $\mathcal{W}_p$. Note that $\mathbf{p}(t)$ is the nominal position of $N$-agent team configuration at any time $t \in \left[t_0,t_f\right]$.
\subsection{Continuum Deformation Optimization}
Our objective is to determine the time-varying factors $\alpha_l(t)$, used in Eq. \eqref{eq:desired_leaders_positions}, for every $l \in \mathcal{W}_1$ and any $t \in \left[t_0, t_f\right]$ by applying quadratic programming, assuming that $\boldsymbol{\beta}_k=\begin{bmatrix} \boldsymbol{\beta}_{i,j,k}\end{bmatrix} \in \mathbb{R}^{|\mathcal{W}_k|\times|\mathcal{W}_{k-1}|}$ is constant for $k \in \{ 2,\cdots,p\}$, where $\left|\cdot\right|$ denotes set cardinality. To this end, we aim to minimize the deviation of the nominal position of $N$-agent team configuration, denoted by $\mathbf{p}(t) = \left[p_x(t), p_y(t), p_z(t)\right]^T$, from the desired position $\mathbf{s}(t) = \left[s_x(t), s_y(t), s_z(t)\right]^T$, at any time $t$, where the nominal position $\mathbf{p}(t)$ is calculated using the following equation:
\begin{equation}
    \label{eq:Wp}
        \mathbf{p}(t) =\begin{bmatrix}
            p_x(t) \\
            p_y(t) \\
            p_z(t)
        \end{bmatrix} = \frac{1}{N_p}\sum_j\mathbf{p}_j(t) \qquad \forall j \in \mathcal{W}_{p} \\
\end{equation}
where $N_p = |\mathcal{W}_p|$ denotes the cardinality of the set $\mathcal{W}_p$. Before further discussion, we first define $\boldsymbol{\alpha}(t)$ as
\begin{equation}
    \centering
    \label{eq:alpha}
    \boldsymbol{\alpha}(t) = \begin{bmatrix}
        \alpha_1(t) & 0 & \dots & 0 \\
        0 & \alpha_2(t) & \dots & 0 \\
        \vdots & \vdots & \ddots & \vdots \\
        0 & 0 & \dots & \alpha_{n_{pl}}(t)
    \end{bmatrix}
\end{equation}
Given material position of leader $l$ as
\begin{equation}
    \mathbf{a}_{l0} = \begin{bmatrix}
        \mathbf{a}_{l0x} &
        \mathbf{a}_{l0y}&
        \mathbf{a}_{l0z} 
    \end{bmatrix}
    ^T,\qquad \forall l\in \mathcal{W}_1, 
    \label{eq:reference_config_1}
\end{equation}
we define the following diagonal matrices:

\begin{equation}
    \centering
    \mathbf{a}_{0x} = \begin{bmatrix}
                    a_{10x} & & \\
                    & \ddots & \\
                    & & a_{n_{pl}0x}
                \end{bmatrix}
                \in \mathbb{R}^{n_{pl}\times n_{pl}},
\end{equation}
\begin{equation}
    \centering
    \mathbf{a}_{0y} = \begin{bmatrix}
                    a_{10y} & & \\
                    & \ddots & \\
                    & & a_{n_{pl}0y}
                \end{bmatrix}
                \in \mathbb{R}^{n_{pl}\times n_{pl}},
\end{equation}
\begin{equation}
    \centering
    \mathbf{a}_{0z} = \begin{bmatrix}
                    a_{10z} & & \\
                    & \ddots & \\
                    & & a_{n_{pl}0z}
                \end{bmatrix}
                \in \mathbb{R}^{n_{pl}\times n_{pl}}.
\end{equation}
Furthermore, we define matrices  $\boldsymbol{\Delta}_x, \boldsymbol{\Delta}_y, \boldsymbol{\Delta}_z$ as follows:
\begin{equation}
    \centering
    \boldsymbol{\Delta}_x = \frac{1}{N_p}\mathbf{1}_{(1 \times N_p)} \boldsymbol{\beta}_p \cdots \boldsymbol{\beta}_2 \mathbf{a}_{0x}\in \mathbb{R}^{1\times n_{pl}},
\end{equation}
\begin{equation}
    \centering
    \boldsymbol{\Delta}_y = \frac{1}{N_p}\mathbf{1}_{(1 \times N_p)} \boldsymbol{\beta}_p \cdots \boldsymbol{\beta}_2 \mathbf{a}_{0y}\in \mathbb{R}^{1\times n_{pl}},
\end{equation}
\begin{equation}
    \centering
    \boldsymbol{\Delta}_z = \frac{1}{N_p}\mathbf{1}_{(1 \times N_p)} \boldsymbol{\beta}_p \cdots \boldsymbol{\beta}_2 \mathbf{a}_{0z}\in \mathbb{R}^{1\times n_{pl}}.   
\end{equation}
Using the definitions above, and by defining the state vector $\mathbf{X}= \left[\alpha_1,\cdots,\alpha_{n_{pl}},s_x, s_y,s_z\right]^T\in \mathbb{R}^{\left(n_{pl}+3\right)\times 1}$, we write individual components of nominal position $\mathbf{p}(t)$ as
\begin{equation}
    \label{eq:pxt}
    p_{x}(t) = \begin{bmatrix}
        \mathbf{\Delta}_x & 1 & 0 & 0
    \end{bmatrix}\mathbf{X}(t) = \mathbf{R}_X \mathbf{X}(t)
\end{equation}
\begin{equation}
    \label{eq:pyt}
    p_{y}(t) = \begin{bmatrix}
        \mathbf{\Delta}_y & 0 & 1 & 0
    \end{bmatrix} \mathbf{X}(t) = \mathbf{R}_Y \mathbf{X}(t)
\end{equation}
\begin{equation}
    \label{eq:pzt}
    p_{z}(t) = \begin{bmatrix}
        \mathbf{\Delta}_z & 0 & 0 & 1
    \end{bmatrix}\mathbf{X}(t) = \mathbf{R}_Z \mathbf{X}(t)
\end{equation}
where  $\mathbf{R}_X, \mathbf{R}_Y, \mathbf{R}_Z$ are of shape $(1,n_{pl}+3)$.

The objective of the quadratic programming problem is assign $\mathbf{X}$ by solving the following optimization problem:
\begin{equation}
\label{eq:quadraticprogramming}
    \minimize_{\mathbf{X}} \hspace{0.1 in} \frac{1}{2}\mathbf{X}^T \mathbf{H} \mathbf{X} + \mathbf{k}^T\mathbf{X}
\end{equation} subject to the following inequality and equality constraints:
\begin{subequations}
\begin{equation}
    \mathbf{A}_{ineq}\mathbf{X} \leq \mathbf{B}_{ineq},
\end{equation}
\begin{equation}
    \mathbf{A}_{eq} \mathbf{X} = \mathbf{B}_{eq}(t),
\end{equation}
\end{subequations}
where
\begin{equation}
    \centering
    \mathbf{A}_{ineq} = \begin{bmatrix*}[r]
        -\mathbf{I}_{(n_{pl}-1) \times (n_{pl}-1)} & \mathbf{0}_{(n_{pl}-1) \times 4} \\
         \mathbf{I}_{(n_{pl}-1) \times (n_{pl}-1)} & \mathbf{0}_{(n_{pl}-1) \times 4}
    \end{bmatrix*}
    \label{eq:Aineq}
\end{equation}
\begin{equation}
    \centering
    \mathbf{B}_{ineq} = \begin{bmatrix*}[r]
        -\alpha_{min} \times \mathbf{1}_{(n_{pl}-1) \times 1} \\
        \alpha_{max} \times \mathbf{1}_{(n_{pl}-1) \times 1} \\
    \end{bmatrix*}
    \label{eq:Bineq}
\end{equation}
\begin{equation}
    \mathbf{A}_{eq} = \begin{bmatrix}
        \mathbf{0}_{4 \times (n_{pl}-1)} & \mathbf{diag}[1,1,1,1]
    \end{bmatrix}
\end{equation}
\begin{equation}
    \mathbf{B}_{eq}(t) = \begin{bmatrix}
        0 \\
        s_x(t) \\
        s_y(t) \\
        s_z(t) 
    \end{bmatrix}
\end{equation}
\begin{equation}
    \centering
    \label{eq:H}
    \mathbf{H} =\zeta \mathbf{I}_{\left(n_{pl}+3\right)\times \left(n_{pl}+3\right)} + \mathbf{R}_X^T\mathbf{R}_X+\mathbf{R}_Y^T\mathbf{R}_Y+\mathbf{R}_Z^T\mathbf{R}_Z
\end{equation}
\begin{equation}
    \centering
    \label{eq:k}
    \mathbf{k}^T = -2(s_x\mathbf{R}_X + s_y\mathbf{R}_Y + s_z\mathbf{R}_z)
\end{equation}
$\mathbf{I}_{(n_{pl}-1) \times (n_{pl}-1)} \in \mathbb{R}^{(n_{pl}-1) \times (n_{pl}-1)} $ is an identity matrix and $\mathbf{0}_{(n_{pl}-1) \times 4} \in \mathbb{R}^{(n_{pl}-1) \times 4}$ is a matrix of zeros in Eq. \eqref{eq:Aineq}. Also,  $\mathbf{1}_{(n_{pl}-1) \times 1} \in\mathbb{R}^{(n_{pl}-1) \times 1}$ represents a vector of ones in Eq. \eqref{eq:Bineq}. In Eq. \eqref{eq:H}, $\zeta>0$ is a small positive number and $\mathbf{I}_{\left(n_{pl}+3\right)\times \left(n_{pl}+3\right)}\in \mathbb{R}^{\left(n_{pl}+3\right)\times \left(n_{pl}+3\right)}$ is an identity matrix. $\alpha_{min}$ and $\alpha_{max}$ are the minimum and maximum values of $\alpha_l(t)$.
\section{Safety Conditions}
\label{sec:safetyconditions}
To ensure safety of the multi-agent team continuum deformation, we provide inter-agent collision avoidance and agent containment conditions by determining $\alpha_{min}$ and $\alpha_{max}$ as discussed in sections \ref{Collision Avoidance} and \ref{Agent Containment} below.
\subsection{Collision Avoidance}\label{Collision Avoidance}
The desired configuration of the agent team is contained by the leading polygon consisting of $n_{pl}-1$ triangles that are defined by the set $\mathcal{T}=\left\{1,\cdots, n_{pl}-1\right\}$. We can express set $\mathcal{V}$ as 
\begin{equation}
    \mathcal{V}=\bigcup_{j\in \mathcal{T}}\mathcal{V}_j,
\end{equation}
where $\mathcal{V}_j$ identifies particular agents enclosed by triangular cell $j\in \mathcal{T}$. Because desired trajectories of the agents are defined by piece-wise affine transformations in Eqs. \eqref{eq:desired_positions_agents} and \eqref{eq:desired_leaders_positions},
the desired position of agent $i\in \mathcal{V}_j$ can be given by
\begin{equation}
    \mathbf{p}_i(t)=\mathbf{Q}_j(t)\mathbf{a}_{i0}+\mathbf{b}_j, \qquad j\in \mathcal{T},~i\in \mathcal{V}_j
\end{equation}
with nonsingular Jacobian matrix $\mathbf{Q}_j\in \mathbb{R}^{3\times 3}$ and rigid-body displacement vector $\mathbf{b}_j\in \mathbb{R}^{3\times 1}$, where $\mathbf{a}_{i0}$ is the reference position of agent $i\in \mathcal{V}_j$. 

We denote
$\lambda_{1,j}(t)>0$, $\lambda_{2,j}(t)>0$, and $\lambda_{3,j}(t)$ as the eigenvalues of pure deformation matrix $\mathbf{U}_j(t)=\left(\mathbf{Q}_j(t)^T\mathbf{Q}_j(t)\right)^{1\over 2}$, and make the following assumptions:
\begin{assumption}
Every agent $i$ can be enclosed by a ball of radius $\epsilon$.
\end{assumption}
\begin{assumption}
Every agent $i$ can stably track the desired trajectory $\mathbf{p}_i(t)$ such that the tracking error does not exceed $\delta$. This implies that
\begin{equation}
    \|\mathbf{r}_i(t)-\mathbf{p}_i(t)\|\leq \delta,\qquad \forall t
\end{equation}
where $\mathbf{r}_i(t)$ and $\mathbf{p}_i(t)$ denote the actual and desired positions of agent $i\in \mathcal{V}$ at time $t$.
\end{assumption}
By evoking the theorem developed in Ref. (\cite{rastgoftar2021safe}), inter-agent collision avoidance is assured  in $j\in \mathcal{T}$, if 
\begin{equation}
   \min\left\{\lambda_{1,j}(t), \lambda_{2,j}(t),\lambda_{3,j}(t)\right\}\geq \dfrac{2\left(\delta+\epsilon\right)}{p_{min,j}},\qquad j\in \mathcal{T},~\forall t,
\end{equation}
where $p_{min,j}$ is the minimum separation distance between every two agents inside $j\in \mathcal{T}$. Then, inter-agent collision avoidance between every two agents in $\mathcal{V}$ is assured, if
\begin{equation}
    \alpha_{min}=\max\limits_{j\in \mathcal{T}}\dfrac{2\left(\delta+\epsilon\right)}{p_{min,j}}
\end{equation}
assigns a lower bound on $\alpha_i(t)$ for every $i\in \mathcal{W}_1$ at any time $t$.

\subsection{Agent Containment}\label{Agent Containment}
We can assure that every agent remains inside a ball of radius $a_{max}$, if 
\begin{equation}
    \alpha_{max}=\dfrac{a_{max}-2\left(\delta+\epsilon\right)}{a_0}
\end{equation}
where $a_0$ is the reference position of boundary leaders' (See Assumption \ref{leadermagassume}). Also, $\delta$ is the control tracking error bound and $\epsilon$ is radius of the circle enclosing every agent $i$, as they were introduced in Section \ref{Collision Avoidance}.
\section{Simulations}
\label{sec:simulations}

\begin{figure*}[ht]
    \centering
    \includegraphics[width=0.94\textwidth]{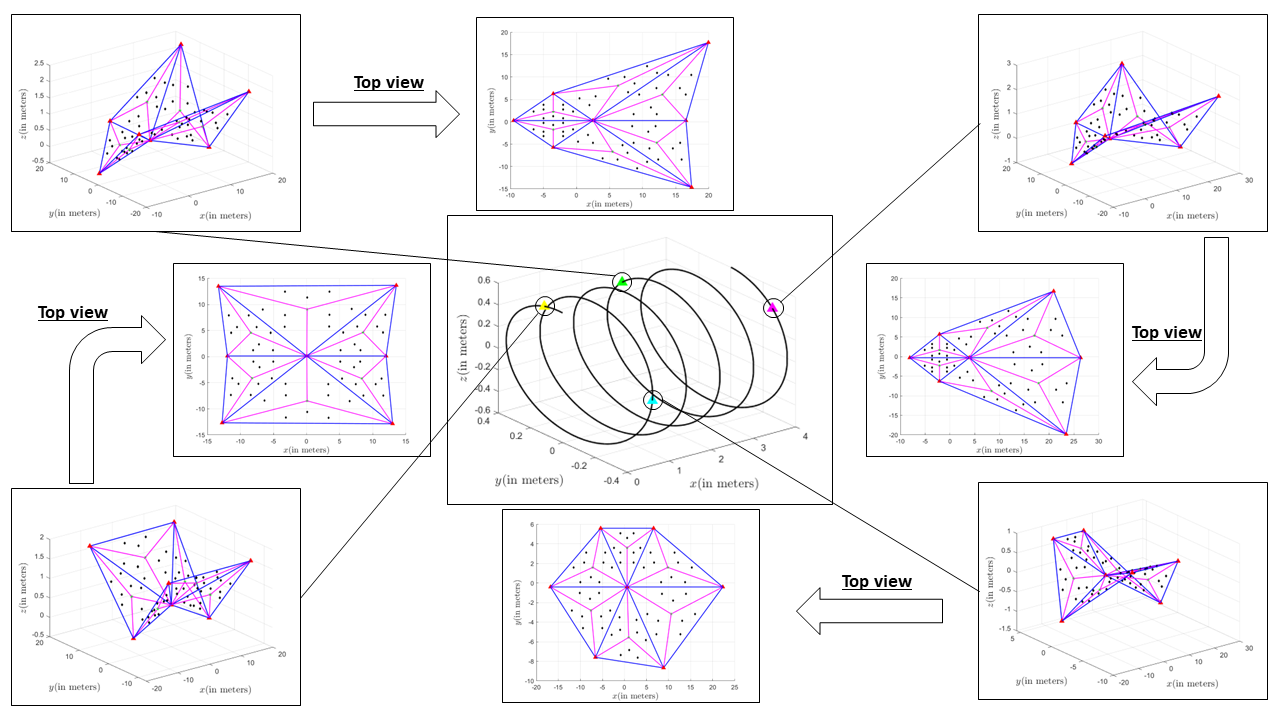}
     \vspace{-0.5cm}
    \caption{Configurations of the MQS at various times while tracking the desired helix trajectory.}
    \label{fig:desired_configs}
\end{figure*}

\begin{figure}[ht]
    \centering
    \subfigure[]{\includegraphics[width=0.48\linewidth]{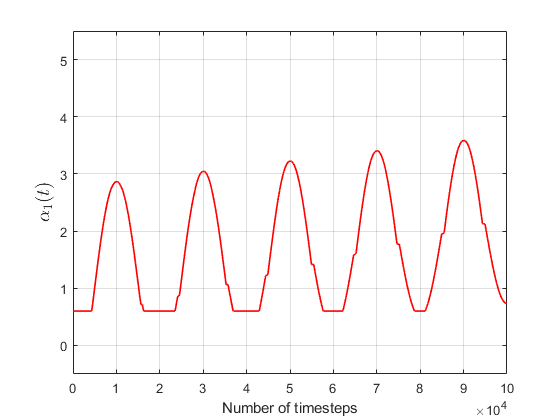}}
    \subfigure[]{\includegraphics[width=0.48\linewidth]{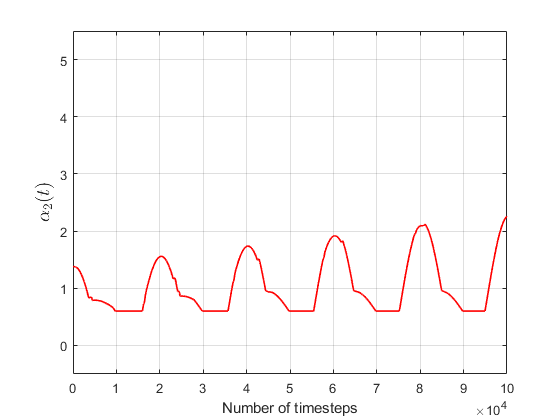}}
    \subfigure[]{\includegraphics[width=0.48\linewidth]{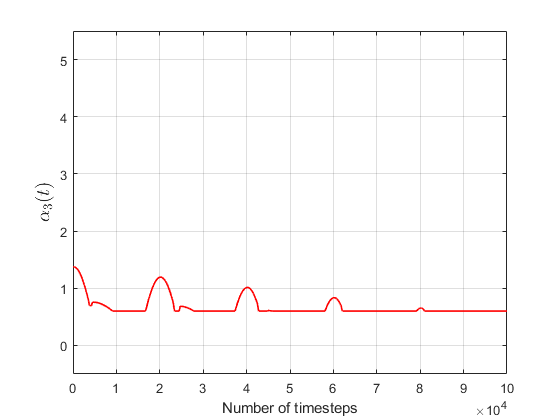}}
    \subfigure[]{\includegraphics[width=0.48\linewidth]{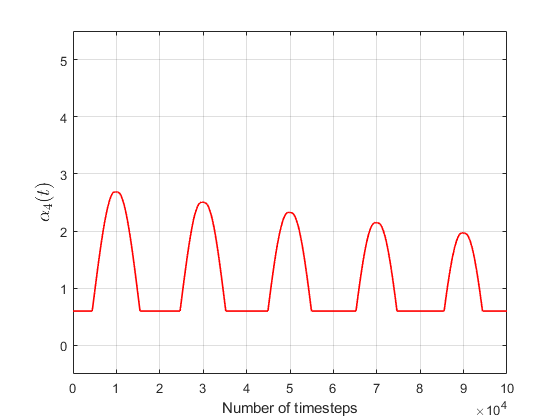}}
    \subfigure[]{\includegraphics[width=0.48\linewidth]{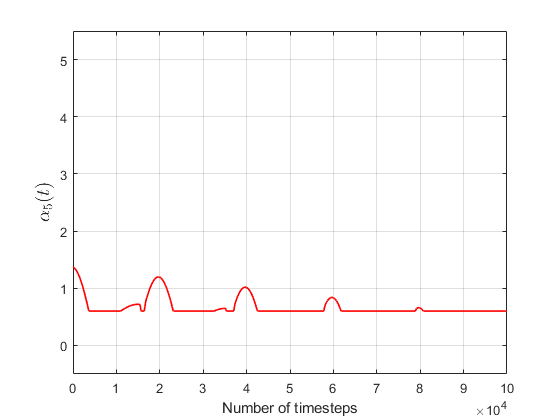}}
    \subfigure[]{\includegraphics[width=0.48\linewidth]{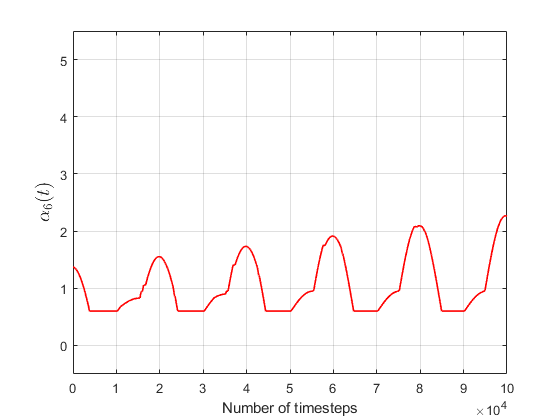}}
    \caption{$\boldsymbol{\alpha}(t)$ for simple helical path}
    \label{fig:all_alphas_helix}
\end{figure}

\begin{figure}[ht]
    \centering
    \subfigure[]{\includegraphics[width=0.32\linewidth]{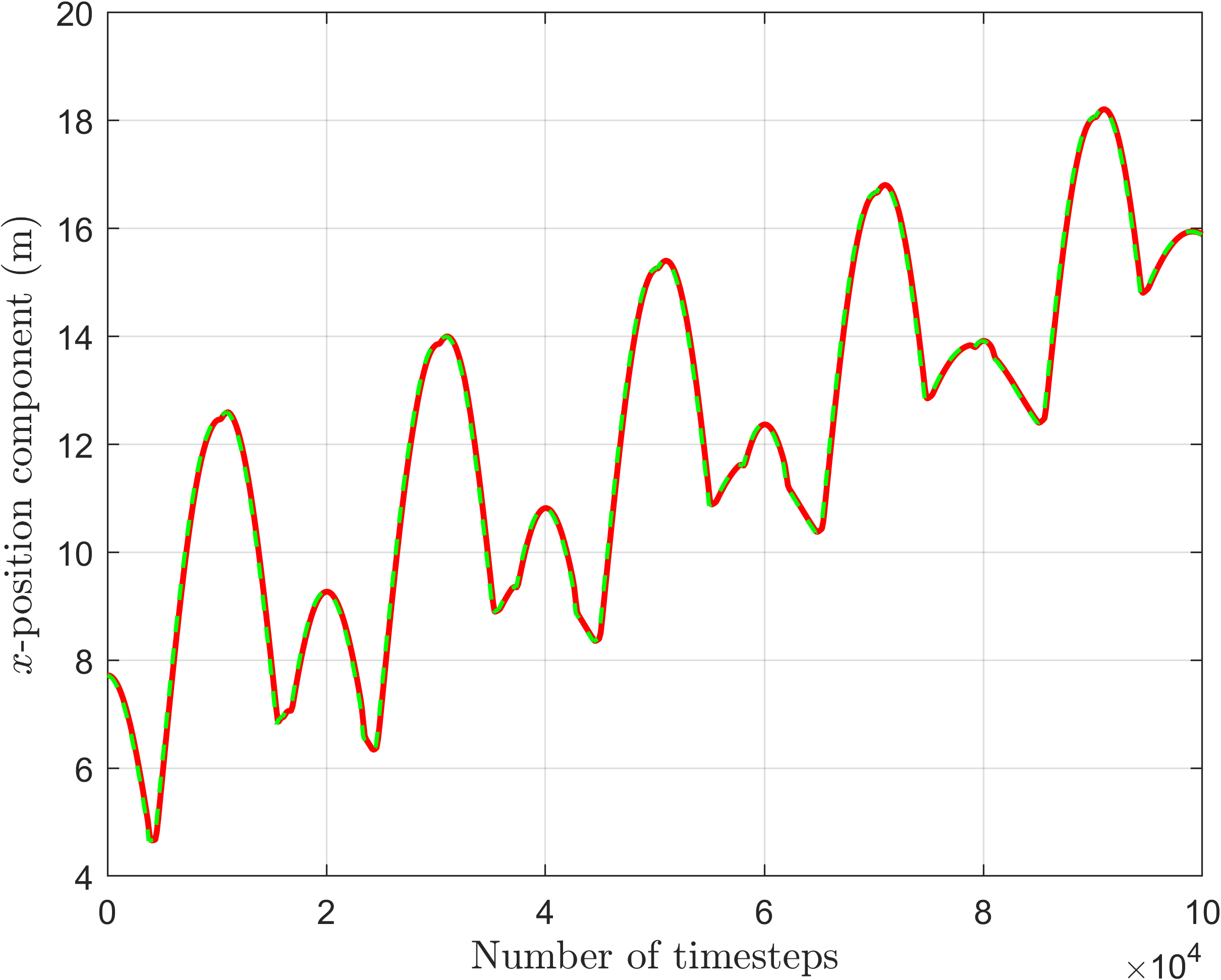}}
    \subfigure[]{\includegraphics[width=0.32\linewidth]{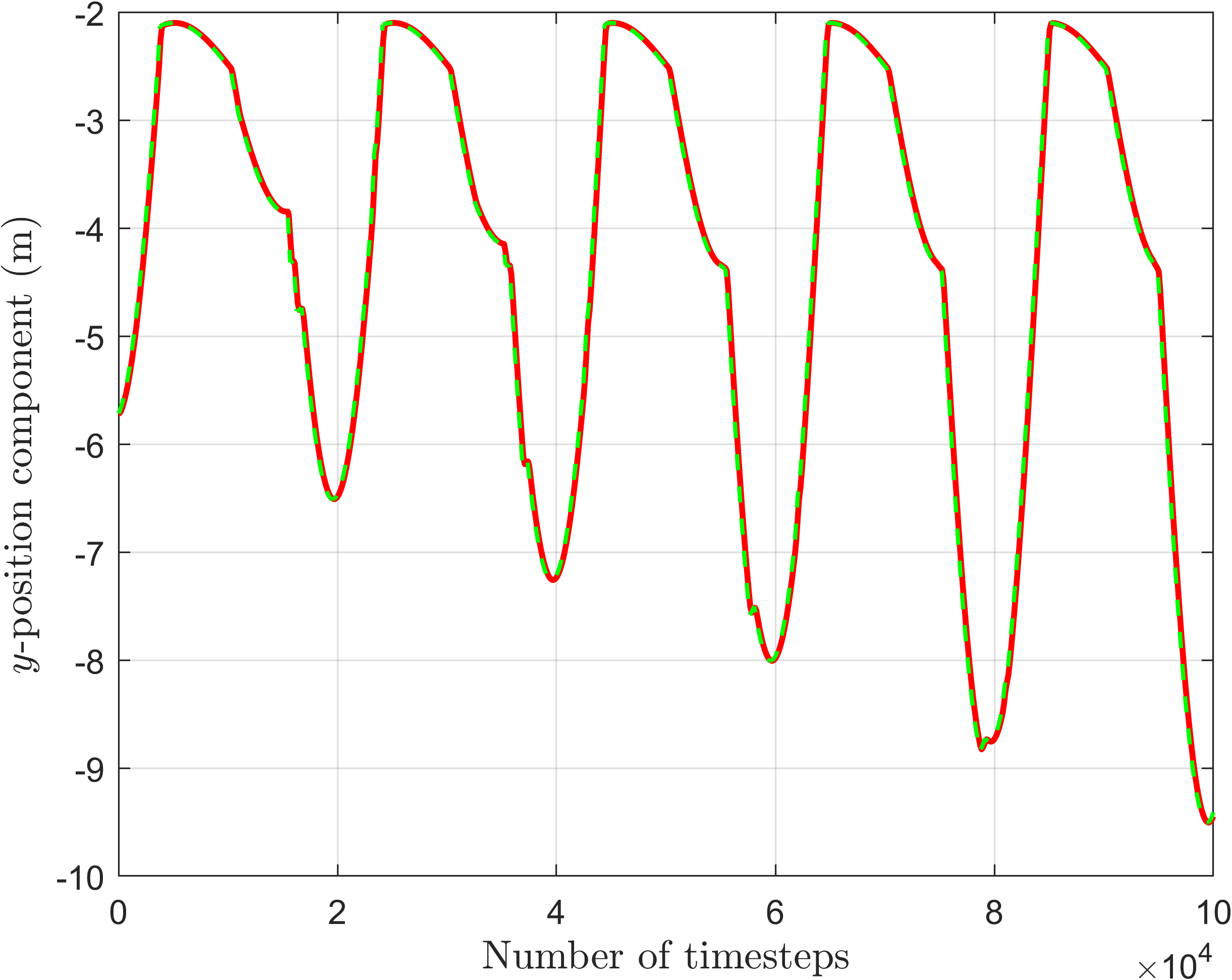}}
    \subfigure[]{\includegraphics[width=0.32\linewidth]{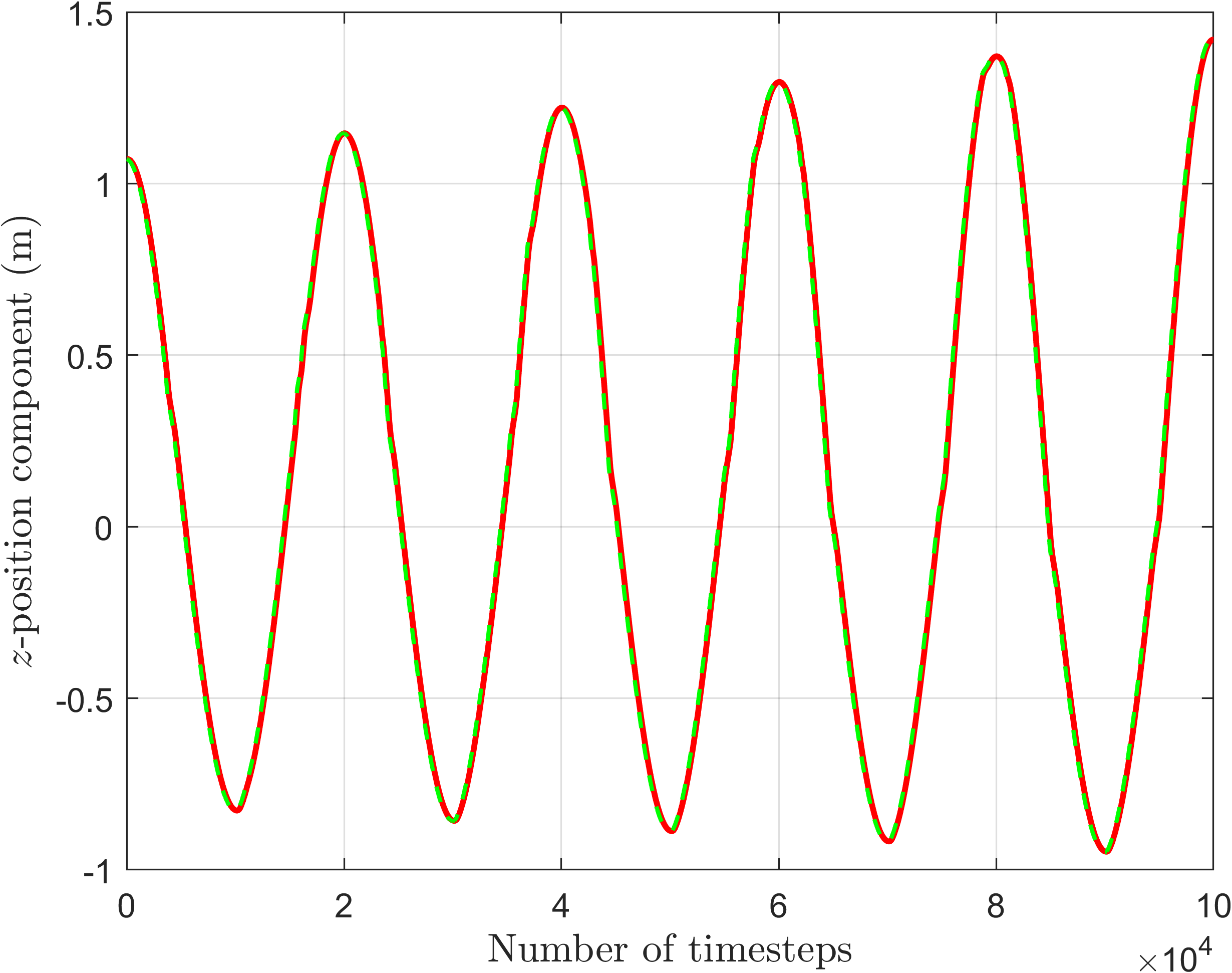}}
    \subfigure[]{\includegraphics[width=0.32\linewidth]{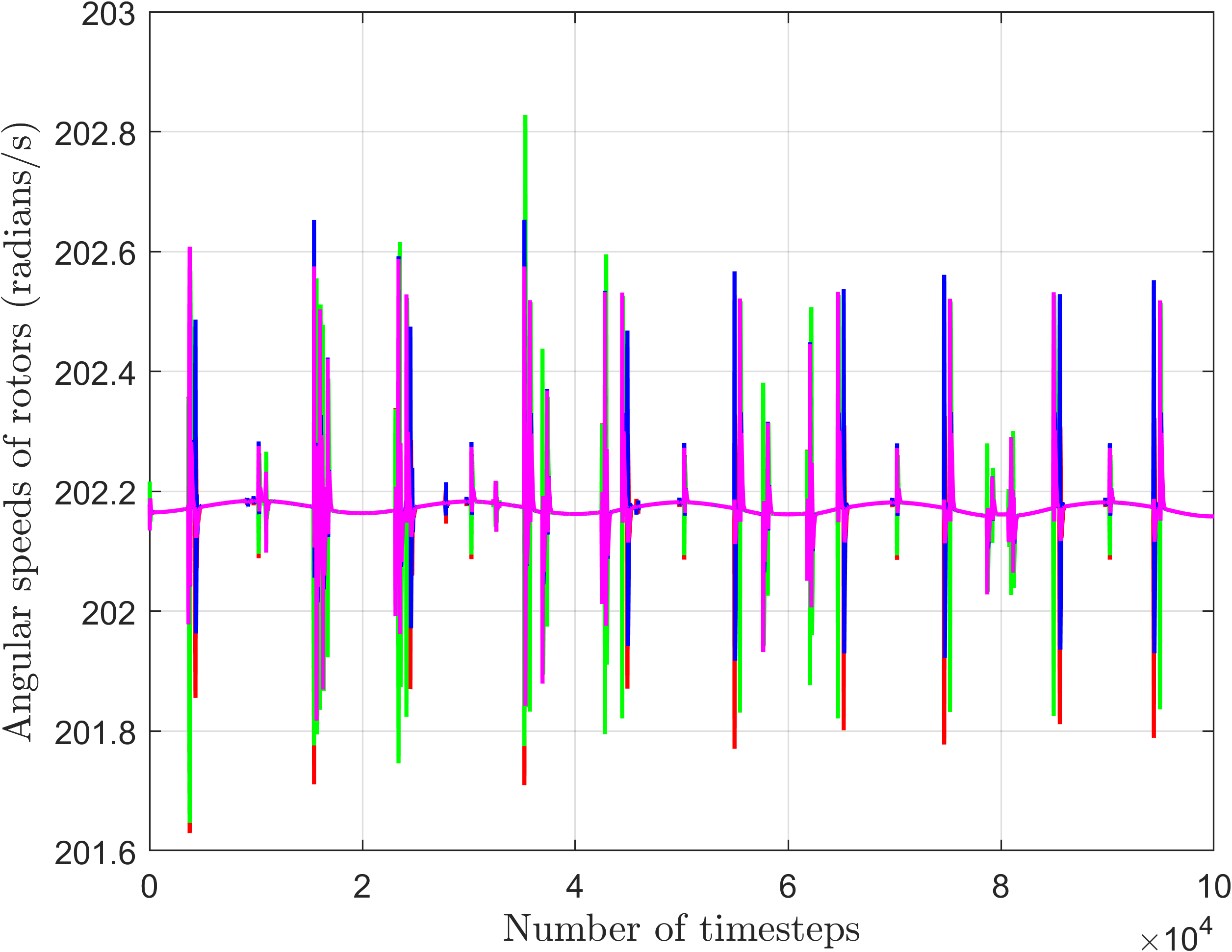}}
    \subfigure[]{\includegraphics[width=0.32\linewidth]{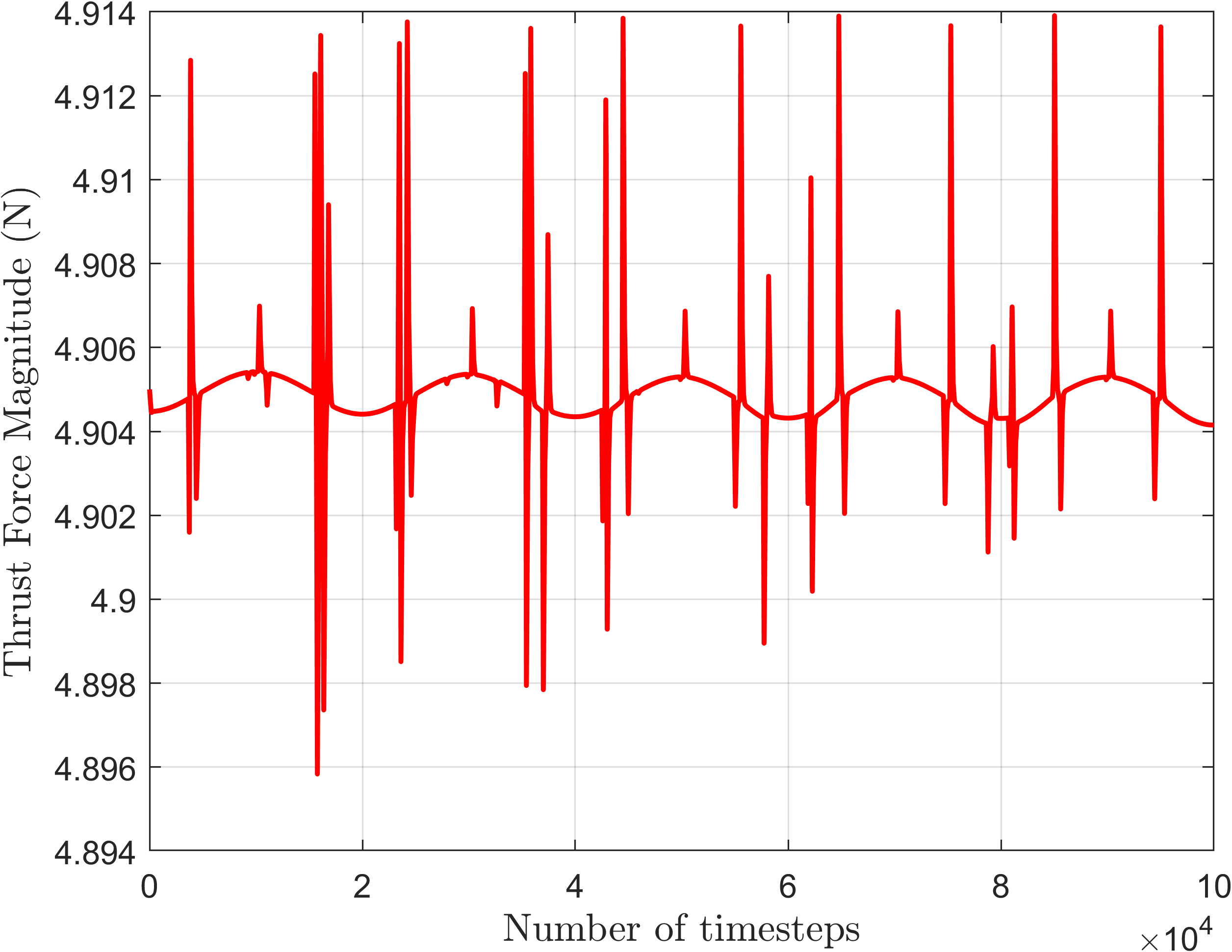}}
     \vspace{-0.5cm}
    \caption{(a) Desired (green) vs Actual (red) $x$-position component, (b) Desired (green) vs Actual (red) $y$-position component, (c) Desired (green) vs Actual (red) $z$-position component, (d) Angular speeds of rotors, (e) Thrust force magnitude for quadcopter agent $67$.}
    \label{fig:all_plots_helix}
\end{figure}

The simulations for this study were conducted on a desktop computer equipped with an Intel i$7$ $11$-th gen CPU, $16$ GB of RAM, and running MATLAB R$2022$b on Ubuntu $20.04$. We investigated the evolution of a multi-quadcopter system (MQS) consisting of $N=67$ quadcopters, which were assumed to have similar size and attributes. The quadcopters' dynamics were modelled based on the work by Rastgoftar (2022) \citep{rastgoftar2022real} and the quadcopter parameters can be found in Gopalakrishnan (2017) \citep{gopalakrishnan2017quadcopter}. 

The quadcopters  are identified by unique index numbers defined by the set $\mathcal{V} = \{1,\cdots,67\}$ with primary leader quadcopters defined by $\mathcal{W}_1 = \{1,\cdots,7\}$. The interior leader quadcopter are given by the set $\mathcal{W}_2 = \{8,\cdots,13\}$ and the followers are denoted uniquely using the set $\mathcal{W}_3 = \{14,\cdots,67 \}$. Therefore, for this simulation we have $p=3$. The minimum and maximum values assigned for the weights $\boldsymbol{\alpha}(t)$ are in the range $\left[0.6, 5.0\right]$. The objective is to determine these weights using Quadratic Programming such that inter-quadcopter collision avoidance is guaranteed (See Section \ref{sec:safetyconditions}) in the MQS. 

For time $t$ in the range $\left[ 0, T\right]$, the large-scale quadcopter team is supposed to move in an obstacle-free environment from an initial configuration, shown in Figure \ref{fig:configuration_plot}, until final time $T$ is reached while tracking the desired helix trajectory. The desired helix trajectory \citep{lee2010geometric} is generated as follows:
        \begin{equation}
        \centering
            \mathbf{s}(t) = \begin{bmatrix}
                s_x(t) \\
                s_y(t) \\
                s_z(t)
            \end{bmatrix} = \begin{bmatrix}
                0.4\omega t \\
                0.4\sin(\pi \omega t) \\
                0.6\cos(\pi \omega t)
        \end{bmatrix}
        \end{equation}
        where $\omega=0.01$ and $T = \SI{1000}{s}$.
The reference (material) positions of the primary leaders are given by
\begin{equation}
\centering
\begin{bmatrix*}[r]
    \mathbf{a}_{10x}&\cdots& \mathbf{a}_{n_{pl}0x}\\
    \mathbf{a}_{10y}&\cdots& \mathbf{a}_{n_{pl}0y}\\
    \mathbf{a}_{10z}&\cdots&\mathbf{a}_{n_{pl}0z}
\end{bmatrix*} = \begin{bmatrix*}[r]
    20 & 10 & -10 & -20 & -10 & 10 & 0 \\
    0 & 10 & 10 & 0 & -10 & -10 & 0 \\
    -1 & 1 & 1 & -1 & 1 & 1 & 0
    \end{bmatrix*}
\end{equation}
In our MATLAB simulations, the state vector $\mathbf{X}$ is defined as $\mathbf{X}= \left[\alpha_1,\alpha_2, \alpha_3, \alpha_4, \alpha_5, \alpha_6, \alpha_7, s_x, s_y, s_z\right]^T$. From equation (\ref{eq:H}), we observe that $\mathbf{H}$ depends on constant weights $\boldsymbol{\beta}_{i,j,k}$ for $k = \{2,\cdots,p\}$ and constant reference (material) configuration $\mathbf{a}_{l0}$. Therefore, $\mathbf{H}$ is constant for all time $t$. However $\mathbf{k}^T$ depends on time-varying desired trajectory $\mathbf{s}(t)$. In Figure \ref{fig:all_alphas_helix}, optimal values of $\boldsymbol{\alpha}(t)$ obtained through quadratic programming have been plotted. Using input-output feedback linearization control approach \citep{rastgoftar2022real}, Figure \ref{fig:all_plots_helix} plots $x$-position component, $y$-position component, $z$-position component, rotors' angular speeds and thrust force magnitude of quadcopter $67 \in \mathcal{W}_3$, respectively. 

\section{Conclusion}
\label{sec:conclusion}


This paper has developed and presented a leader-follower model for large-scale safe and optimal continuum deformation coordination by formulating the optimization as quadratic programming problem. We take inspiration from graphical structure of a neural network to develop a hierarchical optimization framework and obtain time-varying weights over time $t \in [ t_0, t_f]$ using constant reference (material) configuration of the agents as the input. We also provided conditions to assure safety between every two agents while the desired trajectory is tracked by the nominal position of the MAS with minimal deviation. As implied by our simulation results, our approach consisting of $N = 67$ quadcopters is able to track a known target trajectory in $3$-D motion space. Future work in this area can be further extended by making the algorithm more robust as follows: $(1)$ by extending the approach to an obstacle-laden environment integrated with path-planning algorithms such as A$^*$; $(2)$ integrate our previously proposed fluid flow navigation function \citep{UPPALURU2022107960, romano2022quadrotor, emadi2022physics} as an obstacle-avoidance algorithm to account for sudden and abrupt failures in the multi-agent system (MAS); and $(3)$ conduct flight experiments to pass through a different windows of known shape and size.

\bibliography{ifacconf}             

\end{document}